\documentclass[11pt,twoside]{article}  
\usepackage{apn3conf}

\begin{document}   

%
%
%
%

\title{Structured, Dynamo Driven Stellar and Disc Winds}

%
%
%

\author{Brigitta von Rekowski}
\affil{Department of Astronomy \& Space Physics,
       Uppsala University, Box 515, 751 20 Uppsala, Sweden}
\author{Axel Brandenburg}
\affil{NORDITA, Blegdamsvej 17, DK-2100 Copenhagen \O, Denmark}

%
%

\contact{Brigitta von Rekowski}
\email{Brigitta.vonRekowski@astro.uu.se}

%
%
%
%
%

\paindex{von Rekowski, B.}
\aindex{Brandenburg, A.}     

%
%

\authormark{von Rekowski \& Brandenburg}

%
%



\begin{abstract}          
Considerable progress has been made in understanding the hydrodynamics, but only
to a certain extent the magnetohydrodynamics, of shaping bipolar outflows
forming protoplanetary nebulae (PPNs) and planetary nebulae (PNs). In
particular, Blackman et al.\ (2001b,2001a) point out two problems related to the
formation of PNs and PPNs, regarding the formation of multipolar structures and
the origin of the nebulae. They propose a solution by giving a semi-quantitative
physical model which should be investigated by numerical simulations.
\end{abstract}

%
%

\section{Motivation}
{\bf The Shape Problem}.~
All current models involving magnetic fields assume that a radiation driven fast
post-AGB (asymptotic giant branch) wind carries a (frozen-in!) stellar toroidal
magnetic field, travelling into a non-magnetized radiation driven slow AGB wind.
However, it is not clear whether this kind of magnetic shaping can be applied to
nonaxisymmetric {\it multipolar} structures that have been observed by the
Hubble Space Telescope. The multipolar structures suggest that two winds are
interplaying, emanating at the same time from the star and a surrounding {\it
(accretion) disc}, a scenario which has not been considered in previous models.

{\bf The Power Problem}.~
All current models assume that the winds are purely radiation driven. There is
observational evidence that the PN-shaping process begins already in the PPN
phase, because many PPNs have highly collimated fast bipolar outflows. However,
at this early stage of evolution, a post-AGB star is too cool to produce a high
speed (radiation driven!) fast wind with the observed large kinetic luminosity.
Therefore, the {\it origin} of fast PPN winds cannot be radiation, as assumed in
previous models.

{\bf The Proposed Model}.~
Blackman et al.\ (2001b,2001a) propose that large scale dynamos in the post-AGB
star as well as a surrounding PPN accretion disc might drive MHD stellar and
disc winds, so that PPNs/PNs are formed by {\it dynamo driven stellar and disc
winds} rather than radiation driven stellar winds.

{\bf Our Work}.~
We aim, in the future, to verify the proposed semi-quantitative physical model
with {\it global numerical MHD simulations}. As a first step, we present here
results of axisymmetric numerical simulations of three MHD models of a system
comprised of a central object, a surrounding accretion disc and a corona. All
three models have mean-field dynamo action in the disc and do not impose any
external magnetic field in the disc, which is a new approach in the context of
outflows and accretion studies. Model~A assumes an initially non-magnetized
star, whereas Models~B and C include a strong stellar dipolar magnetosphere
and a stellar mean-field dynamo, respectively. Our models are currently applied
to protostellar star--disc systems. However, since we use dimensionless
variables, the models can be rescaled and applied to a range of different
astrophysical objects.
\section{Models and Results}
We solve the continuity equation, the Navier--Stokes equation and the induction
equation. Instead of solving the energy equation, we model a cool dense disc
embedded in a hot rarefied corona by fixing entropy contrasts such that specific
entropy is smaller within the disc and the star, and larger in the corona. By
solving the hydrostatic equilibrium for an initially non-rotating corona, the
corona is pressure-supported, whereas the disc is mainly centrifugally supported
and rotating slightly sub-Keplerian.

{\bf Model~A: Accretion Disc Dynamo}.~
\begin{figure}
\epsscale{.35}
   \plotone{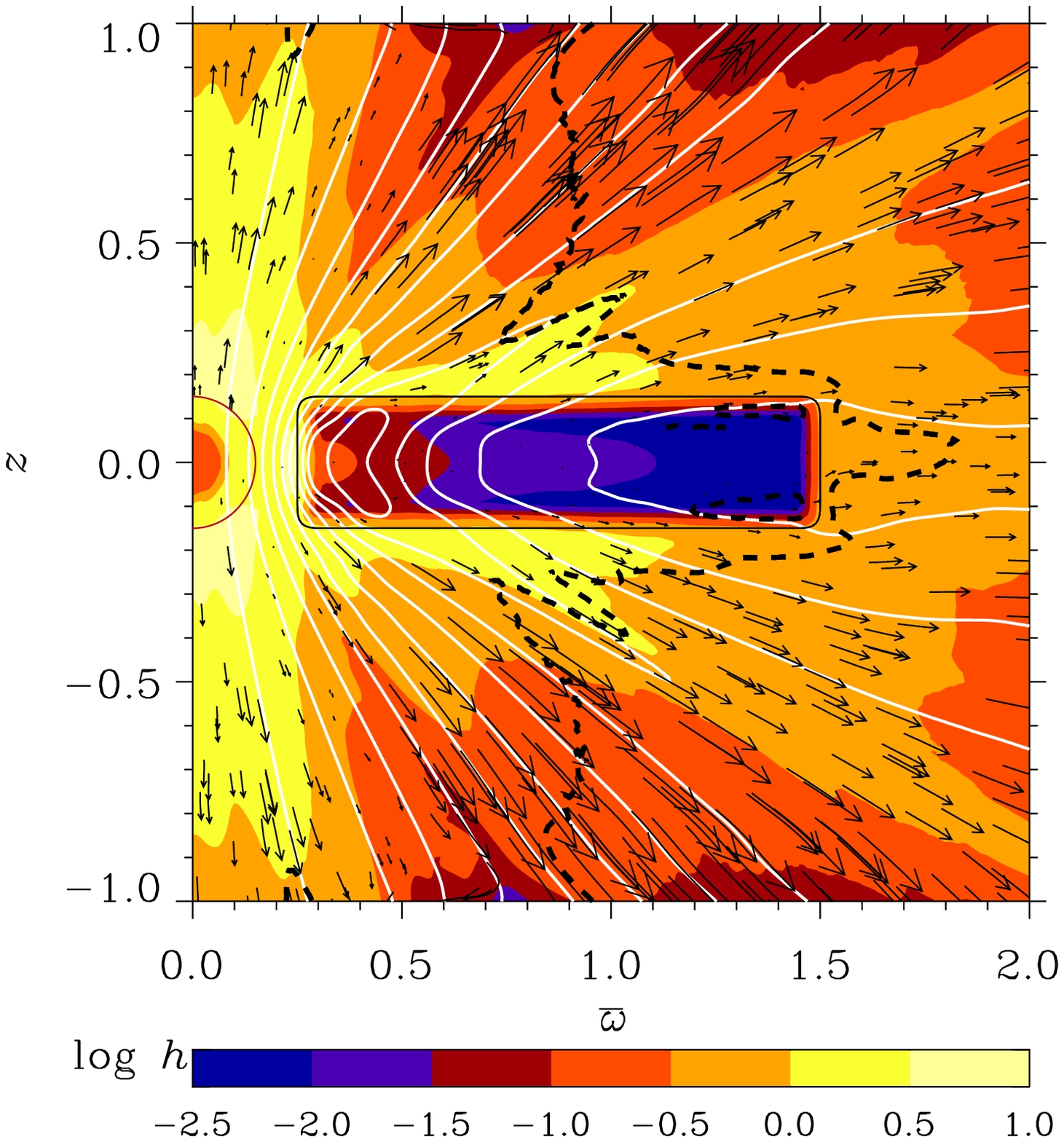}
   \plotone{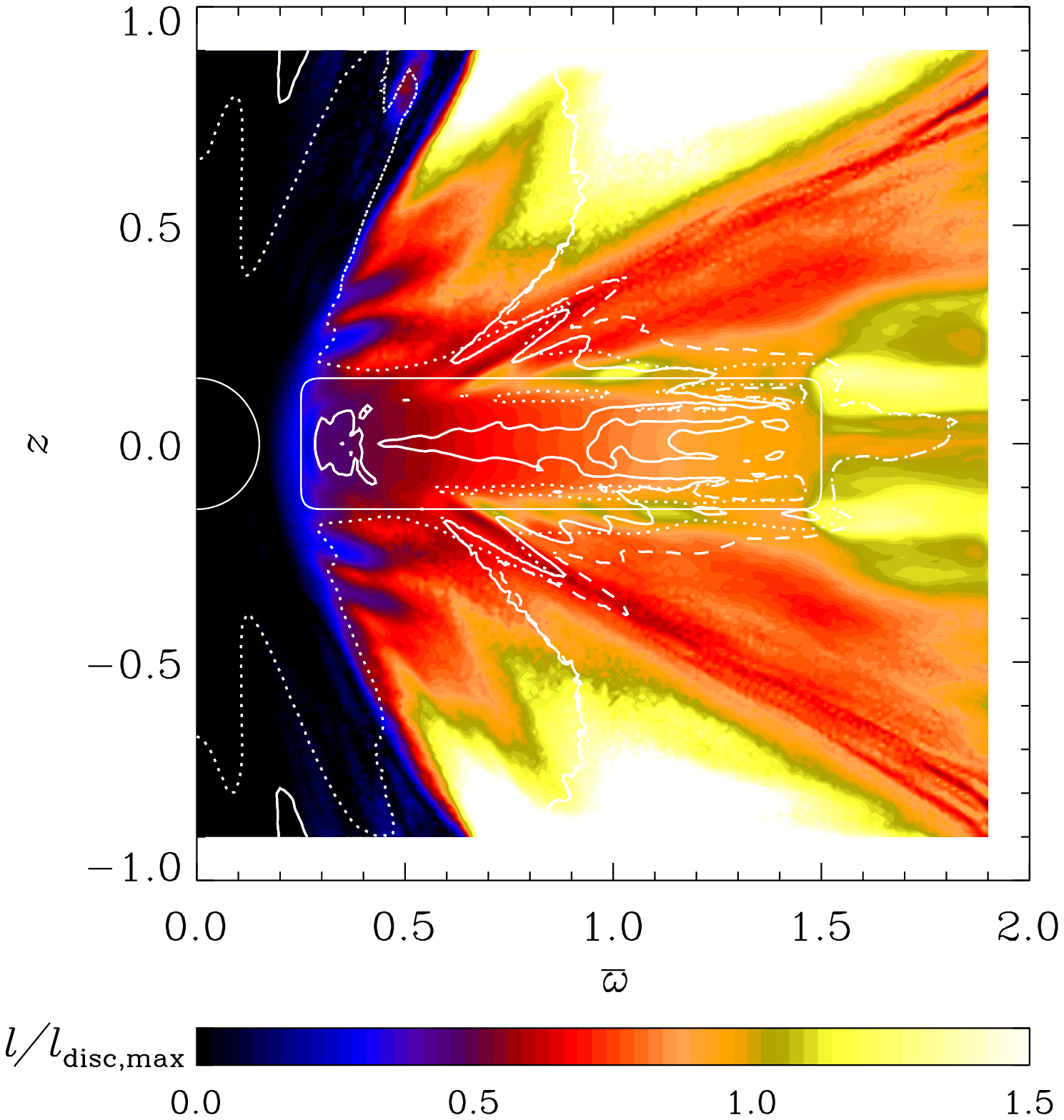}
\caption{Model~A.
Left: Black arrows: poloidal velocity vectors;
white solid lines: poloidal magnetic field lines;
black dashed line: fast magnetosonic surface;
colours: temperature
(specific enthalpy $h$ is directly proportional to temperature $T$,
and $\log_{10}h=(-2,-1,0,1)$ corresponds to $T\approx
(3{\times}10^{3},3{\times}10^{4},3{\times}10^{5},3{\times}10^{6})\,\mbox{K}\,$).
Right: Colours: specific angular momentum $\ell$ (normalized
with the maximum in the disc);
white solid line: Alfv\'en surface;
white dotted line: sonic surface;
white dashed line: fast magnetosonic surface.
Both: Averaged over times $t\approx 897\ {\rm days} \dots t\approx 906\
{\rm days}$; the diffusion time is about $576\ {\rm days}$.
}
\label{Fig1}
\end{figure}
The results of Model~A have been presented in more detail in von Rekowski et
al.\ (2003). As one can see in Fig.~\ref{Fig1} (Left), a clearly structured
outflow develops with a slower, hotter and denser stellar wind, an inner disc
wind (in a conical shell originating from the inner disc edge) that is faster,
cooler and less dense, and an outer disc wind with intermediate values.
The stellar wind is pressure driven (very low angular momentum, see
Fig.~\ref{Fig1} (Right); ratio of the poloidal magneto-centrifugal to pressure
forces is less than $1$ at the stellar surface), whereas the inner disc wind is
magneto-centrifugally accelerated (very high angular momentum; magnetic field
lines at the disc surface are inclined to the rotation axis by at least
$30^\circ$; outflow is highly supersonic but sub-Alfv\'enic: magnetic lever arm
is about $3$; ratio of the poloidal magneto-centrifugal to pressure forces is
larger than $1$ at the disc surface and exceeds $100$ further away), and the
outer disc wind is mostly pressure driven (Alfv\'en surface is close to the disc
surface; ratio of the poloidal magneto-centrifugal to pressure forces is mostly
less than or around $1$ at the disc surface).
The averaged accretion rate is about $10^{-6} M_\odot/{\rm yr}$ and the averaged
disc $+$ stellar wind mass loss rate is about $4 \times 10^{-7}
M_\odot/{\rm yr}$.

{\bf Model~B: Accretion Disc Dynamo \& Stellar Magnetosphere}.~
\begin{figure}
\epsscale{.30}
   \plotone{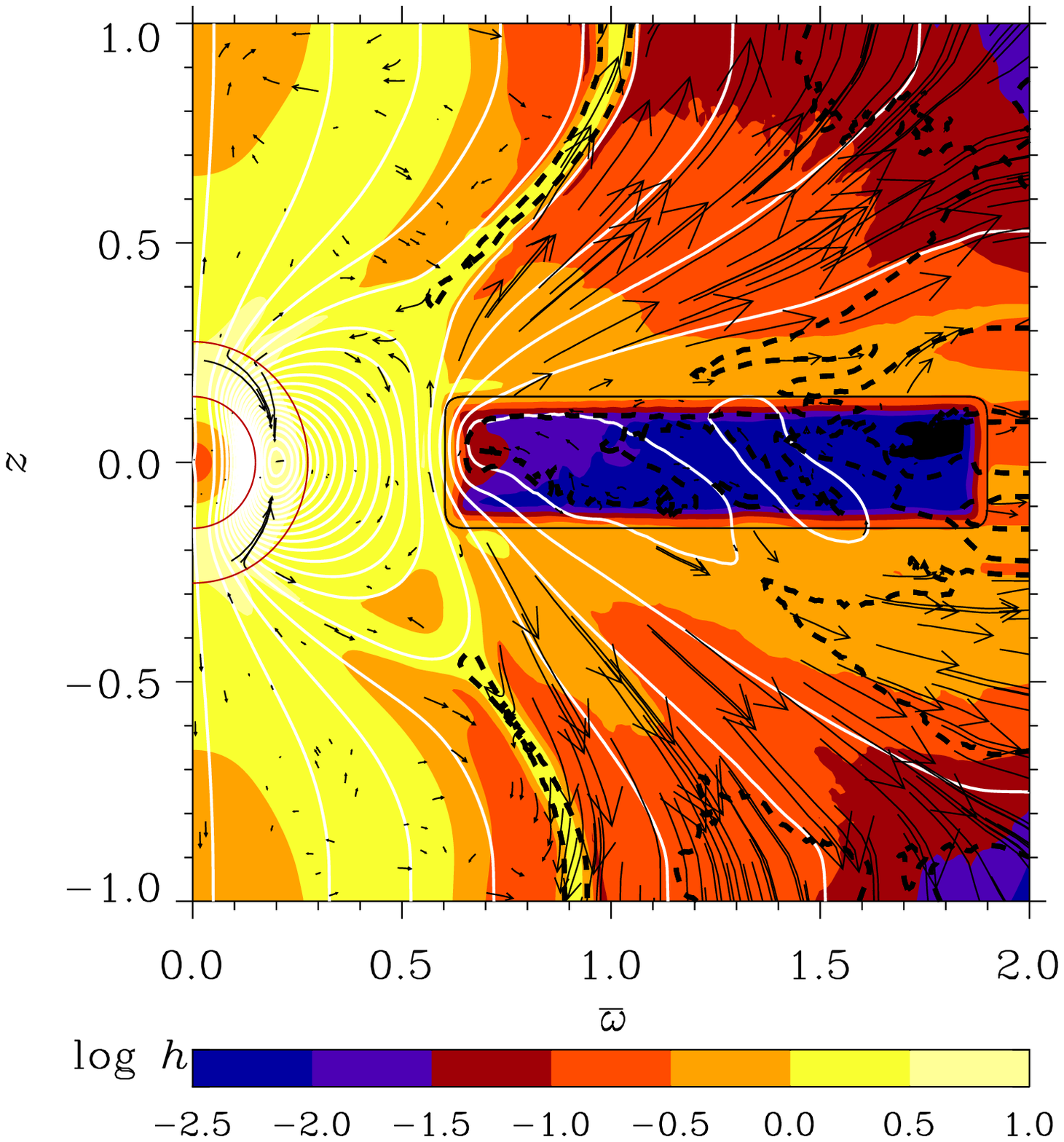}
   \plotone{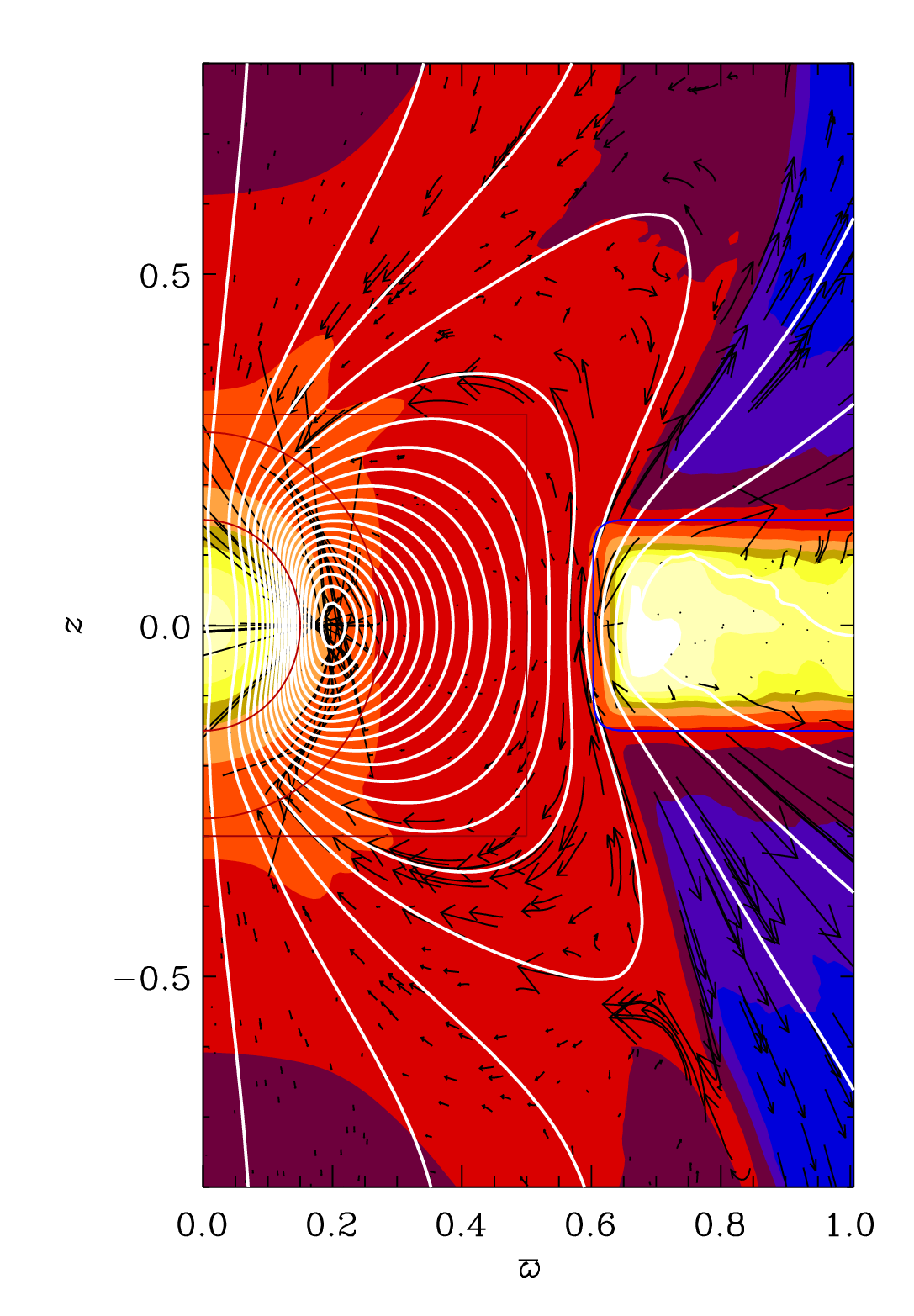}
   \plotone{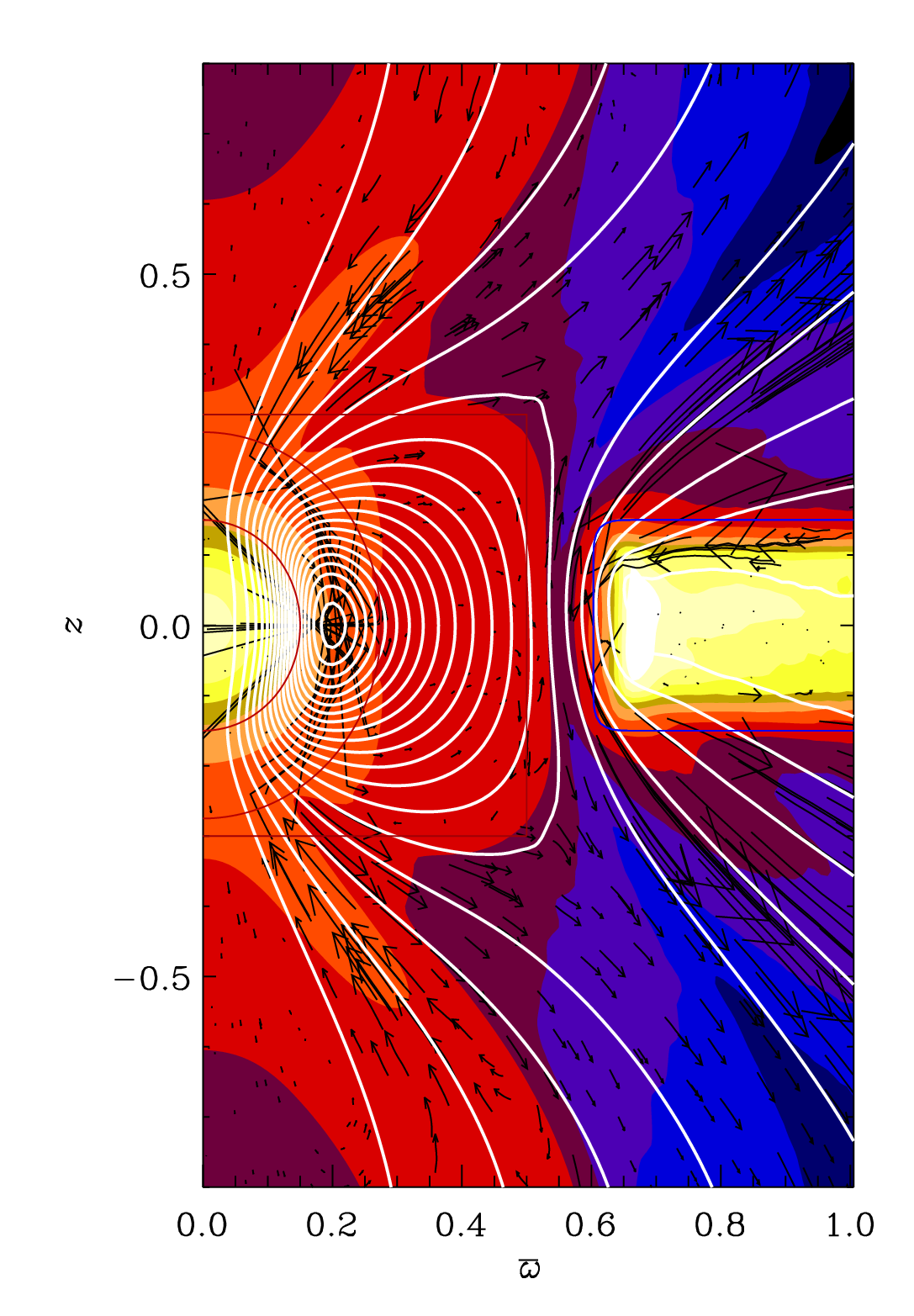}
\caption{Model~B.
Left: Same as in Fig.~\ref{Fig1}, but for Model~B, and
at a time ($t\approx 158\ {\rm days}$) when the star
is about to be magnetically reconnected to the disc.
Black dashed line: Alfv\'en surface.
The stellar surface magnetic field strength is $\approx 5\ {\rm kG}$.
Middle and Right: Colours: density (bright colours indicate high values;
dark colours indicate low values);
white solid lines: poloidal magnetic field lines;
black arrows: azimuthally integrated mass flux density
(except in disc where density is high and vectors would be too long).
The stellar surface magnetic field strength is $\approx 5\ {\rm kG}$.
Middle: At a time ($t\approx 150.5\ {\rm days}$) when the star
is magnetically connected to the disc.
Right: At a time ($t\approx 156.5\ {\rm days}$) when the star
is magnetically disconnected from the disc.
}
\label{Fig2}
\end{figure}
The results of Model~B have been presented in more detail in von Rekowski \&
Brandenburg (2003a). Again, the disc dynamo produces a structured disc wind so
that the overall outflow is similar to Model~A, with coexisting pressure driving
and magneto-centrifugal acceleration. In transition periods when disconnected
stellar and disc field lines are about to reconnect, there is a pressure driven
hot and dense, but relatively fast outflow between the stellar and disc winds
(see Fig.~\ref{Fig2}, Left). The magnetosphere is oscillating with a period of
around 15 to 30 days, changing between two distinct states. When star and disc
are magnetically connected by the magnetosphere, accretion of disc matter onto
the star is along magnetospheric field lines  with peak accretion rates of about
$(1\ldots2.5)\times10^{-8}M_\odot/{\rm yr}$ (see
Fig.~\ref{Fig2}, Middle). When star and disc are magnetically disconnected,
there is no net accretion but matter is lost directly into the outflow (see
Fig.~\ref{Fig2}, Right). In this state, angular momentum transport by the inner
disc wind is clearly enhanced; also the stellar and inner disc wind velocities
are enhanced. The disc wind mass loss rates are also time dependent but not
correlated with the oscillating magnetosphere; they are about one order of
magnitude higher than the peak accretion rates.

{\bf Model~C: Accretion Disc Dynamo \& Stellar Dynamo}.~
\begin{figure}
\epsscale{.35}
   \plotone{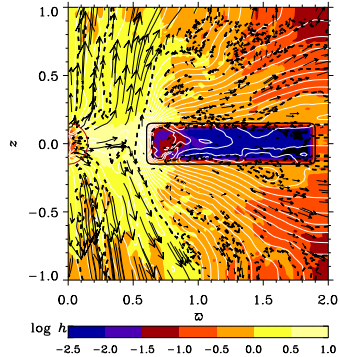}
\caption{Model~C.
Same as in Figs~\ref{Fig1} and \ref{Fig2} (Left), but for Model~C, and
at a time $t\approx 1172\ {\rm days}$.
}
\label{Fig3}
\end{figure}
The results of Model~C are going to be presented in more detail in von Rekowski
\& Brandenburg (2003b). The stellar dynamo generates a magnetic field with
dipolar-type symmetry (the alpha effect is positive in the upper hemisphere).
The main result is that the dynamo in the star produces a fast and collimated
stellar wind (cf.\ Fig.~\ref{Fig3}). The structured disc wind is again due to the
disc dynamo.
\acknowledgments
Use of the supercomputers SGI 3800 in Link\"oping and Linux cluster 1 in Ume\aa\
and of the PPARC supported supercomputers in St Andrews and Leicester is
acknowledged.

%
%
%
%


\end{document}